# Modulation of Ionic Current Rectification in Ultra-Short Conical Nanopores

Long Ma,[1#] Zhongwu Li,[2#] Zhishan Yuan,[3] Chuanzhen Huang,[1] Zuzanna S. Siwy,[4] and Yinghua Qiu[1,5,6]*

1. Key Laboratory of High Efficiency and Clean Mechanical Manufacture of Ministry of Education, National Demonstration Center for Experimental Mechanical Engineering Education, School of Mechanical Engineering, Shandong University, Jinan, 250061, China

2. Jiangsu Key Laboratory for Design and Manufacture of Micro-Nano Biomedical Instruments, School of Mechanical Engineering, Southeast University, Nanjing 211189, China

3. School of Electro-mechanical Engineering, Guangdong University of Technology, Guangzhou, 510006, China

4. Department of Physics and Astronomy, University of California, Irvine, California 92697, United States

5. Advanced Medical Research Institute, Shandong University, Jinan, Shandong, 250012, China

6. Suzhou Research Institute, Shandong University, Suzhou, Jiangsu, 215123, China

[#] These authors contributed equally.
*Corresponding author: yinghua.qiu@sdu.edu.cn

**Abstract:**

Nanopores that exhibit ionic current rectification (ICR) behave like diodes, such that they transport ions more efficiently in one direction than the other. Conical nanopores have been shown to rectify ionic current, but only those with at least 500 nm in length exhibit significant ICR. Here, through the finite element method, we show how ICR of conical nanopores with length below 200 nm can be tuned by controlling individual charged surfaces i.e. inner pore surface ($surface_{inner}$), and exterior pore surfaces on the tip and base side ($surface_{tip}$ and $surface_{base}$). The charged $surface_{inner}$ and $surface_{tip}$ can induce obvious ICR individually, while the effects of the charged $surface_{base}$ on ICR can be ignored. The fully charged $surface_{inner}$ alone could render the nanopore counterion-selective and induces significant ion concentration polarization in the tip region, which causes reverse ICR compared to nanopores with all surface charged. In addition, the direction and degree of rectification can be further tuned by the depth of the charged $surface_{inner}$. When considering the exterior membrane surface only, the charged $surface_{tip}$ causes intra-pore ionic enrichment and depletion under opposite biases which results in significant ICR. Its effective region is within ~40 nm beyond the tip orifice. We also found that individual charged parts of the pore system contributed to ICR in an additive way due to the additive effect on the ion concentration regulation along the pore axis. With various combinations of fully/partially charged $surface_{inner}$ and $surface_{tip}$, diverse ICR ratios from ~2 to ~170 can be achieved. Our findings shed light on the mechanism of ionic current rectification in ultra-short conical nanopores, and provide a useful guide to the design and modification of ultra-short conical nanopores in ionic circuits and nanofluidic sensors.

**Introduction:**

Ionic current rectification (ICR)[1-3] is an important electrokinetic effect, where ionic current for voltages of one polarity is higher than that for voltages of the opposite polarity. ICR has been achieved using pores and membranes as a template. The effect of ICR has been investigated with biological nanopores,[4,5] organic and inorganic artificial micro- and nanopores,[6] as well as heterogeneous porous structures.[1] It was suggested that ionic rectifiers can be applied in a wide variety of fields including ionic circuits,[7] ion sieves,[8] biosensors,[9] desalination,[10] and energy conversion.[11]

ICR arises from voltage-dependent ionic concentrations in a pore or its vicinity. One voltage polarity corresponding to the high conductance state is caused by the accumulation of ions, while the opposite voltage polarity leads to formation of a depletion zone and low ionic conductance. ICR can be induced by asymmetry in nanopore geometry, as done in conically shaped nanopores,[2] as well as nanopores with stepped cross-section.[12] Asymmetric surface properties of nanopores, e.g. asymmetric charge density,[13] wettability[14] and slip length[15] of surfaces along the pore axis can also produce ICR. Nanopores in contact with solutions that differed in conductivity achieved by concentration gradients,[16] and viscosity gradients[17,18] were reported to exhibit current rectification as well. Finally, the presence of additional charged surfaces near one end of the pore could also induce ICR.[19] In all cases mentioned above, surface charges at solid-liquid interfaces played an important role in the observed asymmetric ion transport due to local ionic selectivity and electroosmotic flow (EOF) [20].

The majority of rectifying nanopores were created based on nanopores whose length exceeded 500 nm. It is because that zones with depleted or enhanced ionic concentrations have a finite length and need to be confined within the pore for the appearance of ICR.[1-3] Short nanopores, however, offer many advantages compared to long structures, such as larger ionic permeation coefficient,[21] and higher sensitivity in bio-sensing.[22] Thus, achieving significant ICR in short nanopores may be of great importance in improving the efficiency of energy conversion and desalination, as well as the sensitivity of biosensors.

Examples of short rectifying nanopores have already been reported: many of the systems were based on bipolar nanopores which contained a junction between positively and negatively charged zones of pore walls. Relative high ICR ratios could be obtained due to the opposite ionic selectivity of both pore ends to cations and anions. Using advanced nanofabrication approach, Yan *et al.*[23] prepared bipolar silica isoporous membranes with a total thickness of ~150 nm, which exhibited a rectification ratio ~3 in KCl solutions. Yang *et al.*[24] fabricated nanopores in a 50 nm-thick SiN membrane with ~30 nm gold deposition on one side, which showed an ICR ratio ~5 in 0.1 M KF solution due to the induced bipolar charge distribution on gold surfaces. Through finite element method, Xie and co-workers[25] explored current rectification in bipolar cylindrical pores with 10 nm diameter. When the pore length was shorter than 200 nm, ICR ratios less than ~10 were observed. With molecular dynamics simulations, Luan *et al.*[26] studied current-voltage characteristics of triangular nanopores in bilayer hexagonal boron nitride membranes with positive and negative charges on the pore rim of different layers. ICR ratios above 5 could be achieved when the boundary length of nanopores was less than 1 nm.

Different from bipolar nanopores, in nanopores with uniform surface charges, ICR typically requires asymmetric shape such as a conical geometry, a nanoscale tip diameter, and a low bulk concentration.[2, 27] When electric fields are applied, ionic enrichment or depletion can form inside the pore, which induces the high or low conductance of the nanopore, i.e. the "on" or "off" states of the ionic diode behavior.[2] In these uniformly charged conical pores, ICR was also found to depend on the pore length,[28] such that rectification in short pores, i.e. less than 200 nm in length, was rarely observed. In nanopores prepared by dielectric breakdown in 30 nm-thick $SiO_2$ pyramidal membranes, the ICR ratio was less than 2.[29] In another system, 55 nm-long truncated pyramidal silicon nanopores with a fixed cone angle at 54.7° did not rectify the current for the tip diameters between 4.7 to 57.5 nm in KCl solutions 1 to 1000 mM.[30] In the simulations conducted by Zhang *et al.*,[31] the ICR ratio was ~2 in a strongly charged conical pore with 200 nm in length.

With the development of nanofabrication techniques, short conical nanopores could be prepared conveniently through various methods in SiN and $SiO_2$ membranes, such as track-etching,[32] anisotropic chemical etching,[30, 33] dielectric breakdown,[29, 34] as well as combination of electron beam lithography and reactive ion etching.[35] These nanopores have provided a versatile platform for the study of nanofluidics. In order to reveal the mechanism underlying current-voltage characteristics of short conical nanopores and find effective ways to create strong ionic current rectifiers using ultra-short conical nanopores, we provide understanding how ICR ratios can be tuned in conical nanopore systems with length below 200 nm. Taking advantages of the simulations, we probed the influence of individual charged surfaces of the nanopore. The charged inner surface and exterior surface on the tip side were found to play important roles in current rectifying.[28, 36, 37] We have identified conditions that lead to preparation of 100 nm long ionic rectifiers with ICR ratios varying from ~2 to ~170.

**Simulation Method:**

3D Simulations were conducted with COMSOL Multiphysics. Influences of surface charges, electric double layers (EDLs) and electroosmotic flow on ionic transport were considered through coupled Poisson-Nernst-Planck (PNP) and Navier-Stokes (NS) equations.[18, 38] Steady-state ionic current and ion distributions were obtained at the room temperature 298 K. Figure 1 and Figure S1 show the simulation scheme. 8 types of models with different charged parts of the pore walls have been considered, i.e. the cases with all charged surfaces (ACS), no charged surfaces (NCS), charged inner surface (ICS), charged both exterior surfaces (ECS), charged exterior surface on the tip side ($ECS_t$), charged exterior surface on the base side ($ECS_b$), charged inner surface and exterior surface on the tip side (IECS$_t$), as well as charged inner surface and exterior surface on the base side (IECS$_b$). The subscripts t and b mean tip and base, respectively. More simulation details are provided in Supporting Information (Figure S2, Tables S1 and S2).

**Results and Discussions:**

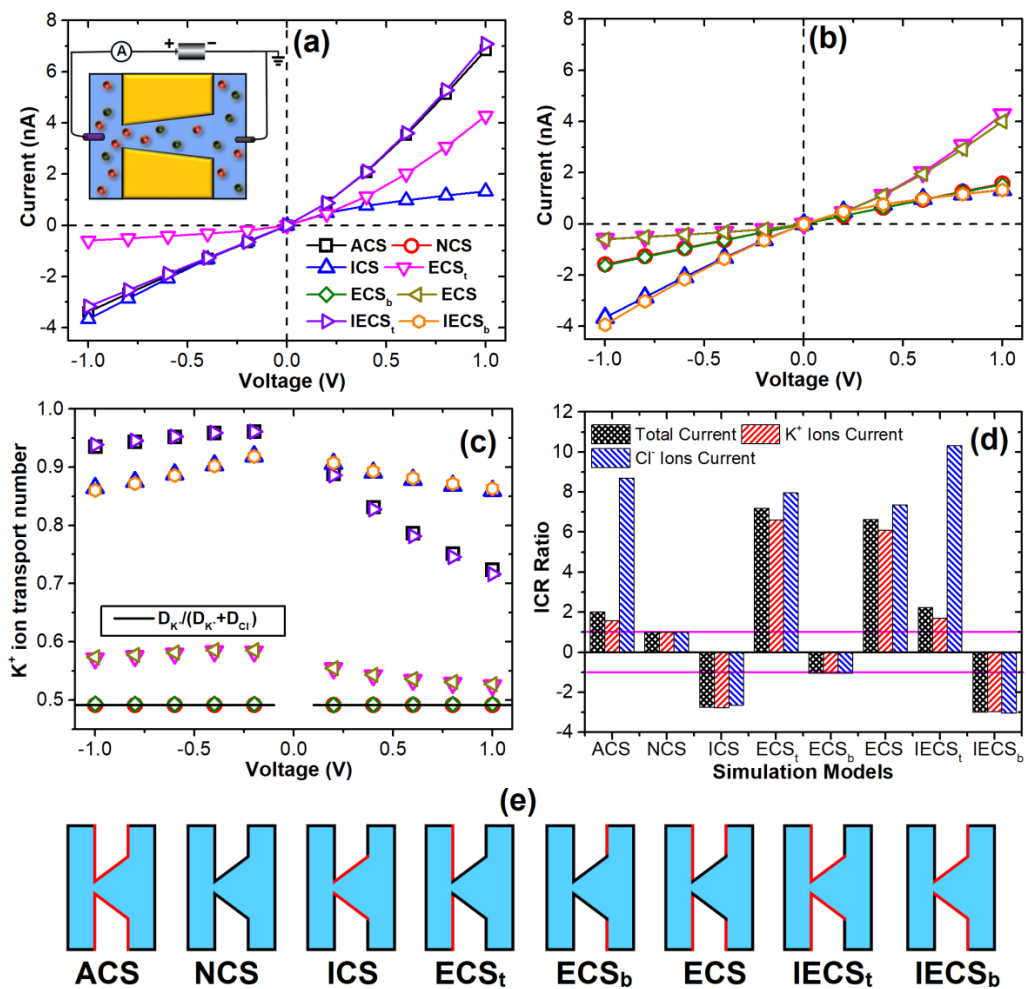

Figure 1 Ionic current behaviors in 8 conical nanopores with different arrangement of surface charges. The pores were 100 nm in length, the tip and base opening was 2 and 20 nm in radius, respectively. (a) and (b) Current-voltage (IV) curves. The inset in (a) shows the setup of voltage polarity in simulations. (c) K$^+$ ions transport number. (d) ICR ratios across conical nanopores at ±1 V calculated from (±) $|I|_{larger} / |I|_{smaller}$. Positive and negative values mean larger ionic currents were observed at 1 V and −1 V, respectively. (e) Scheme of 8 simulations models. Charged surfaces are shown in red. In simulations, 0.1 M KCl aqueous solution was selected. Surface charge density of −0.08 C/m$^2$ was used for charged surfaces.

**Weak ICR in the ACS model – nanopores with homogeneously charged surfaces**. Figure 1 shows ionic current characteristics of 8 models of conical nanopores with different charged surfaces. We define the ICR direction as forward when higher ionic currents correspond to K$^+$ ions moving from the tip to the base of conical nanopores, i.e. in our electrode configuration with the working electrode placed at the tip side, positive currents are higher than negative currents. If the positive currents are lower than negative currents, we call the ICR reverse.

Among the 8 nanopore systems, the largest current at voltages of both polarities occurred in the ACS model, due to the contribution from accumulated counterions in the EDLs near charged surfaces. However, because of the weak direction-dependent flow characteristics of the main current carriers (K$^+$ ions) through the nanopore, the ICR ratio is only ~2, which agrees well with simulation results from Pietschmann *et al.*[28] with 187-nm-long conical nanopores.

For the ACS model, we have additionally considered how ionic transport depends on different nanopore properties such as surface charge density, tip size, and half-cone angle, as well as simulation conditions including salt type and concentration, and the presence of EOF, as shown in Figure S3 and S4. Some of the conditions considered were also reported in earlier reports.[36, 39-41] We found that ICR of 100 nm-long nanopores in the ACS model was surprisingly insensitive to the surface charge density in the range of −0.08 to −0.24 C/m$^2$. In contrast, in long conical nanopores, the increase of charge density led to a significant improvement of ICR due to stronger enhancement and depletion of ions at voltages of opposite polarities.[39] Even more surprising was the observation that the tip size and bulk ionic concentrations do not significantly change ICR of short ACS nanopores either.[20] Similar to earlier results, adjusting the half-cone angle,[40] and diffusion coefficient of counterions[36] in short pores did not improve ICR ratios. The influence of EOF[41] on the current rectifying was not obvious either, because the strong confinement of the nanopore hinders the fluid flow. Finally, we calculated ICR for nanopores with different lengths. The modeling revealed that with the pore length increasing from 50 to 300 nm, the ICR ratio enhanced from ~1.5 to ~3 (Figure S5). Our simulations suggested that short conical nanopores with homogenous surface charges provide only weak rectification with ICR ratios not exceeding 3.

In the remaining models with one or multiple charged surfaces, the effects of different charged parts of the conical nanopore on ionic current have been investigated separately.

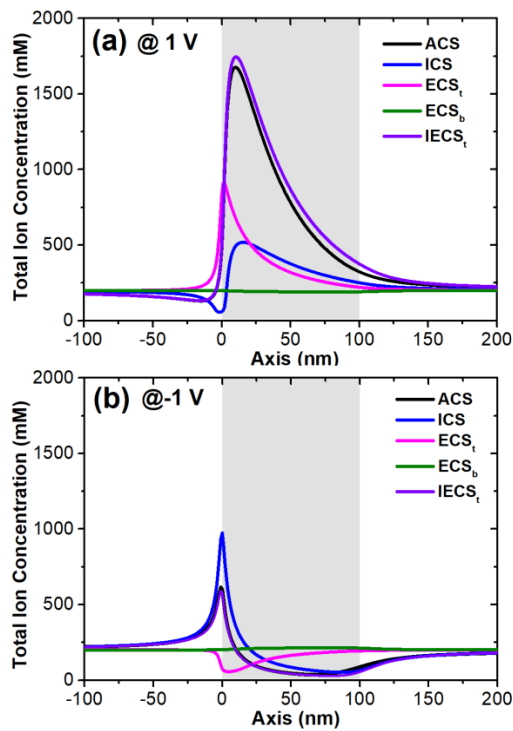

Figure 2 Distributions of total ion concentrations along the pore axis in 5 types of simulation models at 1 V (a) and −1 V (b). Pore regions are shown in light grey. Ion concentration distributions in the rest cases are shown in Figure S6.

**ICR with varied directions in the ICS model – pores with charged inner walls**. In the ICS case, the high ionic selectivity to $K^+$ ions causes obvious ion concentration polarization (ICP) across the tip region. As shown in Figure 2, when the pore tip is in contact with a positive electrode, a clear depletion zone is formed in the solution outside the tip opening while ionic enrichment occurs inside of the pore. Under the opposite voltage polarity, ionic concentrations are enhanced at the tip entrance, while the pore volume is depleted of ions. Since the regulation of ionic concentrations at the tip opening has a larger influence on ionic current than that inside the pore, this nanopore exhibits ICR of direction that is opposite to the ACS model as well as long conical nanopores.[42] In addition, without charged exterior surfaces, the ICR ratio can be enhanced to ~−3, thus higher than ICR in the ACS system.

Similar phenomena of reverse ICR have also been reported in conical nanopores with only charged inner surface.[31, 37, 43]. As shown by Zhang *et al.*[31] and Yan *et al.*[42], in long nanopores with only charged inner surfaces, the intra-pore ion enrichment and depletion dominate the modulation in ionic concentrations, which produce forward ICR. However, with the nanopore length decreasing, the stronger electric field strength causes more obvious ICP in front of the tip mouth which finally determines the current rectifying into the opposite direction.

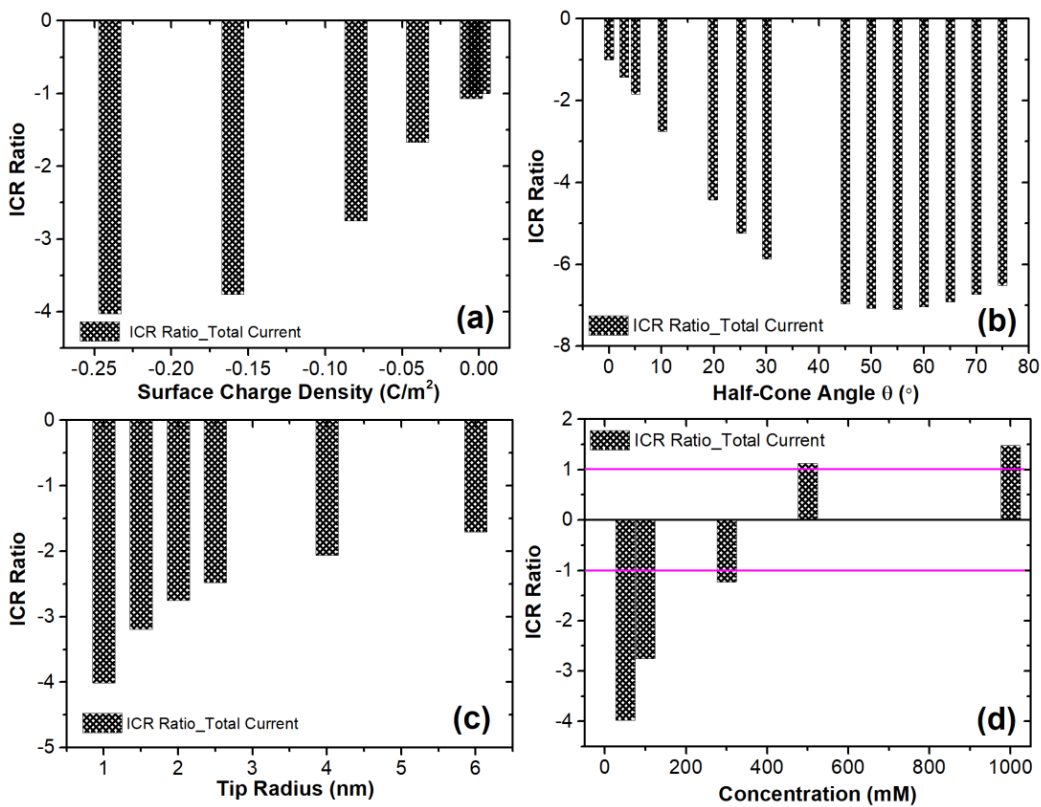

Figure 3 ICR ratios of total ionic current under ± 1 V from the ICS model. Different nanopore properties were considered, such as surface charge density (a), half-cone angle (b), and tip radius (c), as well as a different simulation condition, i.e. ionic concentration (d).

We also considered the effect of nanopore properties and simulation conditions on reverse ICR (Figure S7). From Figure 3, with the surface charge density changing from 0 to −0.24 C/m$^2$, the ICR ratio is enhanced from −1 to ~−4. This observation could be explained by a more significant modulation of ionic concentrations by ICP, leading to formation of a deeper depletion zoon at the tip region under positive voltages. The half-cone angle also influences ICR: increasing the angle from 0 to 55⁰ modulates ICR ratio effectively from −1 to ~−7. According to ionic concentration distributions along the pore axis in Figure S8, as the half-cone angle increases, the effective pore length decreases. Consequently, a stronger electric field at the tip enhances the effect of ICP.[40, 44] However, with the further increase of the half-cone angle beyond ~55 ⁰, the smaller charged area on the inner wall, which can provide strong electrostatic interaction to mobile ions, induces weaker ICP and lower ICR ratios.

The tip opening diameter was also important for tuning ICR in nanopore with charged inner walls. Narrower pores are more cation selective and more affected by ICP than wider nanopores. When changing the bulk ionic concentration from 50 to 1000 mM (Figure 3d), we observed an interesting effect of reversing ICR direction. As the ionic strength increases, more counterions accumulate near the charged surface and screen the charges. Both ICP at the tip opening and intra-pore ion enrichment/depletion become weaker (Figure S9). When the bulk concentration reaches 500 mM or above, ICP is nearly negligible and the ionic enrichment/depletion inside the nanopore starts to dominate the current rectifying leading to forward ICR. Note that our results have a similar trend to the earlier report showing a 1 μm-in-length conical nanopore that changed the rectification direction at ~20 mM.[37] In Figure S10, effects of salt types and EOF on ICR have also been studied, which does not affect ionic transport in a significant manner.

36

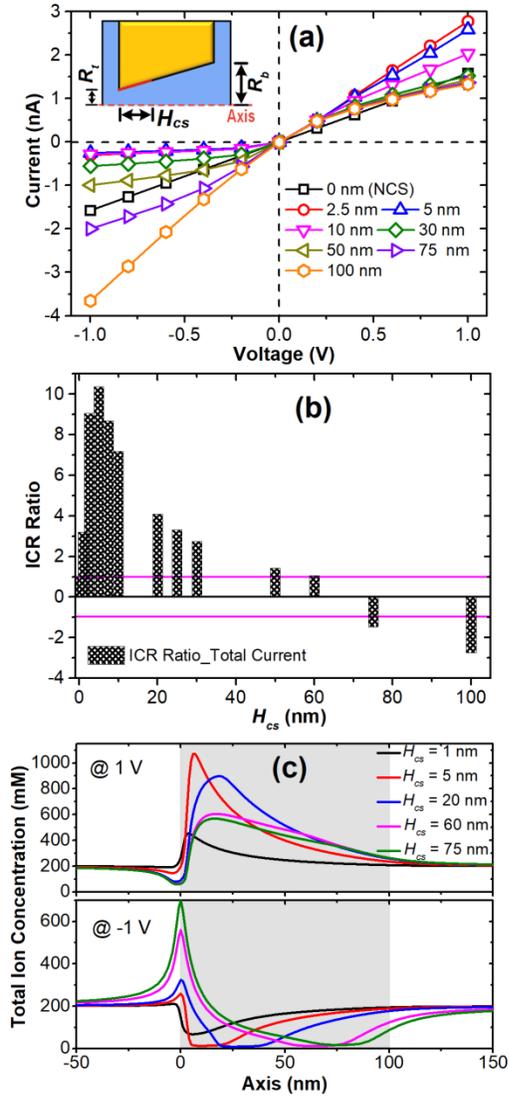

Figure 4 Ionic current characteristics from the ICS model. The depth of the charged zone $H_{cs}$ was varied from 0 to 100 nm. (a) IV curves. The inset shows the scheme of $H_{cs}$. (b) ICR ratios of current at ±1 V in the pores with different $H_{cs}$. (c) Distributions of the total ion concentration along the pore axis in the cases with different $H_{cs}$. Pore regions are shown in light grey.

Transport properties of conical nanopores are strongly affected by charge state of the tip region of the pore.[9, 11] In long pores, placing a junction between charged and uncharged zone close to the tip can induce significant changes of current-voltage curves.[45] Here, we probed how presence of such junction can enhance ICR in short conical nanopores. A junction with its charged zone was considered at the tip which depth was defined as $H_{cs}$ (Figure 4a) varying between 0 and 100 nm. It had a complex influence on the current characteristics. When the inner surface gets charged even with a little area at the very tip, the nanopore starts to show current rectification. With $H_{cs}$ increasing to 10 nm, ionic current values at positive biases decrease, while those at negative biases show only small variations. As $H_{cs}$ increases further, the current at positive biases gradually saturates, and those at negative biases start increasing and achieve their maximum values at $H_{cs}$ =100 nm, i.e. the inner surface is fully charged.

The resulting ICR for different $H_{cs}$ is shown in Figure 4b. For low values of $H_{cs}$, ICR was forward, with maximum ratio at $H_{cs}$ = 5 nm. This short charged region already led to significant modulations of ionic concentrations in the pore, with minimal ICP. With the further increase of $H_{cs}$, the ICR ratio started to decrease, with a turning point at ~60 nm where an ICR ratio close to 1 was predicted. With $H_{cs}$ > 60 nm, the nanopore rectified in the reverse direction. For a conical nanopore that contains a junction between charged and uncharged zones, all ICR properties can be explained by the interplay between the shape induced concentration modulations and ICP.[31, 42]

In the 100-nm-long conical nanopore, the length of the charged zone $H_{cs}$ that leads to the highest ICR is 5 nm. The dependence of ICR on the length of the charged zone was earlier considered in long conical nanopores where an optimal magnitude of $H_{cs}$ was predicted as $\sim (R_t/R_b)L$, when $R_t \ll R_b$.[9] When applied here, the formula predicted the highest ICR to occur for $H_{cs}$ = ~10 nm, a value that is quite close to our results $H_{cs}$ = 5 nm. The dependence of the optimal $H_{cs}$ on the pore half-opening angle was also investigated, Figure S11. With the base size increasing, the optimal charged depth becomes smaller, which is consistent with Vlassiouk *et al.*'s prediction. In our system, the ICR ratio reaches maximum values at the charged depth which provides the most significant intra-pore ion enrichment and depletion. A unipolar conical nanopore with a neutral tip region was also studied. From Figure S12, different from earlier simulations,[9, 46] obvious reverse ICR was achieved that we attributed to the ion concentration regulation by ICP at the tip region.

**Significant ICR in the ECS$_t$ model – nanopores with only charged exterior membrane surface on the tip side.** In Figure 1, significant ICR of ~7 can be obtained with the ECS$_t$ model. This system exhibits forward ICR, similar to the ACS case. The total current value in the high conductance state is ~60% of those in the ACS model, and the nanopore is only weakly K$^+$ selective at a level of ~60%. However, due to the strong confinement of the tip opening, the ionic concentrations in the pore can still be influenced by the external voltage, such that positive voltages cause enhancement of ionic concentrations while negative voltages cause formation of a depletion zone.

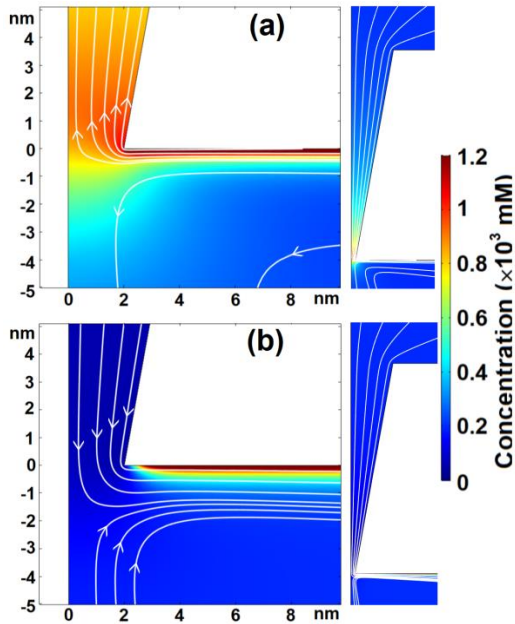

Figure 5 Distributions of total ion concentration and electroosmotic flow near the tip opening from the ECS$_t$ model, at 1 V (a) and −1 V (b). Color map shows the ion concentration. White lines and arrows represent the electroosmotic flow direction.

The effect of surface charges on exterior pore walls has often been ignored in modeling of long conical nanopores.[9, 47, 48] While some recent studies showed that charged exterior surfaces had significant effects on the ionic current through nanopores.[37, 49] In short conical nanopores, the influence of charged exterior surface on the tip side on current rectifying has seldom been investigated. In order to understand the mechanism of ICR in the $ECS_t$ model, we investigated not only ionic concentrations but also characteristics of fluid flow near the tip opening.[18] From Figure 5a, when the pore conducts more current at 1 V, dense streamlines of EOF appear near the charged exterior wall, that point to the inside of the nanopore. The EOF is caused by the directional movement of counterions in EDLs. At positive voltages, the electric field drives $K^+$ ions in the EDLs parallel to the exterior surface into the tip mouth, which accumulate inside the pore. Similar EOF streamlines are also found in the ACS case (Figure S13). We think this phenomenon occurs in all conical nanopores with a charged exterior surface on the tip side. Under positive voltages, the EDLs outside the pore act as an ion pool which provide extra ions needed to form intra-pore ion enrichment. In Figure 5b and S13b, the streamlines point to the outside of the nanopore, which means $K^+$ ions move out of the nanopore at −1 V. Due to electrostatic repulsion between surface charges and $Cl^-$ ions, entrance of $Cl^-$ ions to the pore is hindered. Consequently, ionic depletion is formed under negative voltages.

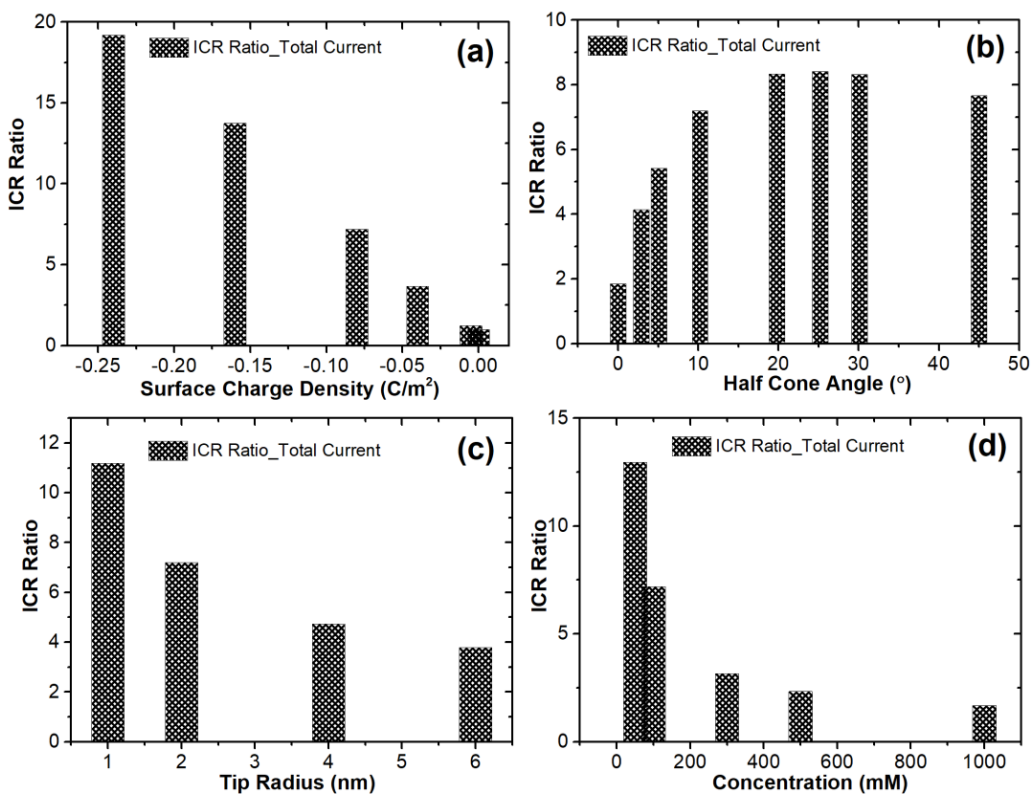

Figure 6 ICR ratios of total ionic current under ± 1 V in the $ECS_t$ model. Different conditions were considered, such as surface charge density (a), half-cone angle (b), and tip radius (c), as well as a different simulation condition, i.e. ionic concentration (d).

For the $ECS_t$ model, nanopore geometrical properties and simulation conditions were considered. In Figure 6, higher surface charge densities attract more counterions in the EDLs which form more distinct ionic enrichment in the pore (Figure S14) and higher ICR ratios (IV curves are shown in Figure S15) are observed. The half-cone angle affects the effective thickness of the nanopore which regulates the potential distribution along the pore axis (Figure S16).[40] With the increase of the half-cone angle, the effective thickness becomes smaller and larger ionic current is obtained. When the angle varies from 0⁰ to ~25⁰, due to the shorter effective pore length and relative stronger electric filed strength in the tip region, more counterions can enter the nanopore, which enhances ICR. However, as the half-cone angle becomes larger than ~25⁰, the ICR ratio decreases again due to weaker enrichment and depletion of ions in the nanopore (Figure S17). These are caused by the promoted ionic diffusion away from the pore inside to the reservoir or into the nanopore from reservoir through the large base opening under positive and negative biases, respectively (Figure S18). The decrease of the tip opening enhances the influence of surface charges on ionic transport that leads to increased ICR.

The bulk ionic concentration also has a strong effect on the ICR ratio in pores with a charged membrane surface. In more diluted solutions, Debye length becomes larger.[20] With the higher degree of EDLs overlap, more obvious ICR can be obtained due to the enhanced intra-pore ion enrichment/depletion (Figure S19). From Figure S20, ICR shows only a weak dependence on the salt type and the consideration of EOF.

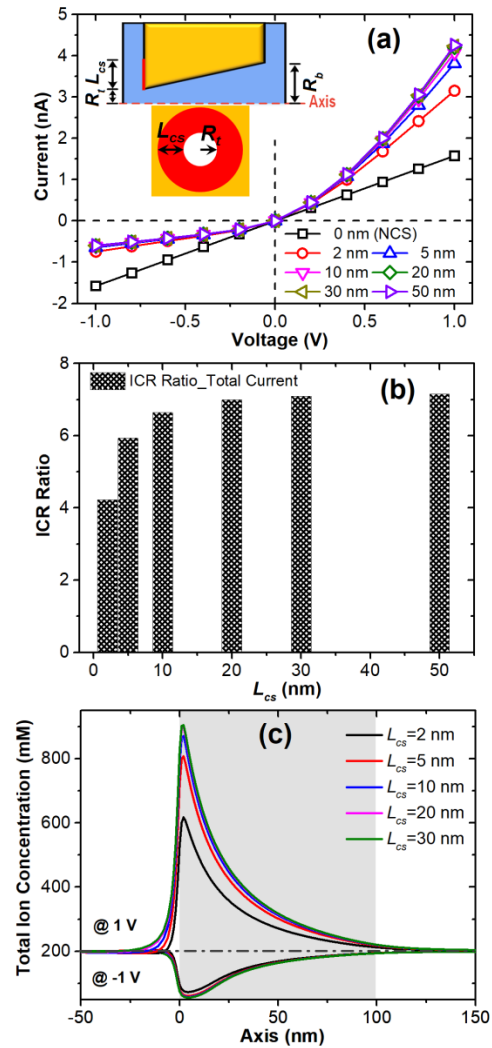

Figure 7 Ionic current characteristics from the $ECS_t$ model. The charged exterior surface length ($L_{cs}$) varying from 2 to 50 nm. (a) Current-voltage curves. (b) ICR ratios of current at ±1 V in the pores with different $L_{cs}$. (c) Distributions of total ion concentrations along the pore axis at ±1 V in the cases with different $L_{cs}$.

Results in Figure 5 suggest that the intra-pore ion enrichment and depletion depends on the charged exterior surface on the tip side. We were also interested in understanding whether the length of the charged exterior surface around the tip opening can influence ionic current.[49] As shown in Figure 7, due to the nanoscale tip opening, even a 2 nm-wide-ring region around the pore can induce obvious ICR of ~4. With the charged width ($L_{cs}$) increasing to 20 nm, the difference between IV curves obtained in this case and the one with $L_{cs}$=~5 μm becomes less than 1%, which can be explained by the almost identical distributions of ion concentration along the axis in both cases (Figure 7c). 20 nm is the minimum effective charged length to control ionic behaviors in the conical nanopore. When $L_{cs}$ is larger than 20 nm, ionic current reaches stable values, which produce constant ICR ratios and ionic selectivity. In the cases with the opposite charged pattern to that shown in the inset of Figure 7a, i.e. a neutral area near the pore opening but a charged region far away, with the neutral width increasing, forward ICR disappears (Figure S21).

In Figure S22, we investigated the influence of nanopore properties and simulation conditions on the minimum length of the charged exterior surface on the tip side. Through analysis of the difference between current values obtained in the nanopores with $L_{cs}$=20 nm and those in the cases with $L_{cs}$=~5 μm, the minimum effective charged length can be controlled by the nanopore properties and electrolyte concentration. With the increase of surface charge density, half-cone angle, or tip size, as well as the decrease of electrolyte concentration, the needed minimum charged length becomes longer. By controlling the current difference obtained in the cases with different $L_{cs}$ from that with $L_{cs}$=~5 μm under 3%, we found the minimum effective charged length is ~40 nm with the surface charge density increasing from −0.005 to −0.24 C/m$^2$, half-cone angle ranging from 0 to 45⁰, tip diameter changing from 2 to 12 nm, or ionic concentration varying from 50 to 1000 mM, as shown in Figure S23.

**Ignored ICR in the ECS$_b$ model** – **nanopores with only charged exterior membrane surface on the base side.** The charged exterior surface near the base end has the weakest influence on the ionic current. Consequently, this model predicts a linear IV curve and no ionic selectivity (Figure 1). Due to the large opening diameter of the base end, EDLs induced by surface charges outside the pore cannot influence the current value and ionic selectivity effectively.

**Modulation of ICR in nanopores with combinations of individual charged parts of nanopores.** From our results, current rectification in ultra-short conical nanopores can occur due to two different mechanisms: intra-pore concentration modulation and ICP in the tip region. For the nanopores with the charged exterior surface on the tip side (ECS$_t$ model, case i) or charged inner pore surface with 5 nm in length near the tip opening (denote as ICS$_t$ model, case ii, Figure 8), forward ICR can stem from induced intra-pore ion enrichment/depletion. While, the 95 nm-long charged inner pore surface near the base end (denote as ICS$_b$ model, case iii, Figure 8) results in reverse current rectifying due to the concentration regulation by ICP at the tip region. Here, in order to probe whether ICR properties can be additive we considered the case (iii) with positive surface charges of 0.08 C/m$^2$ such that this conical pore also rectified in the forward direction.

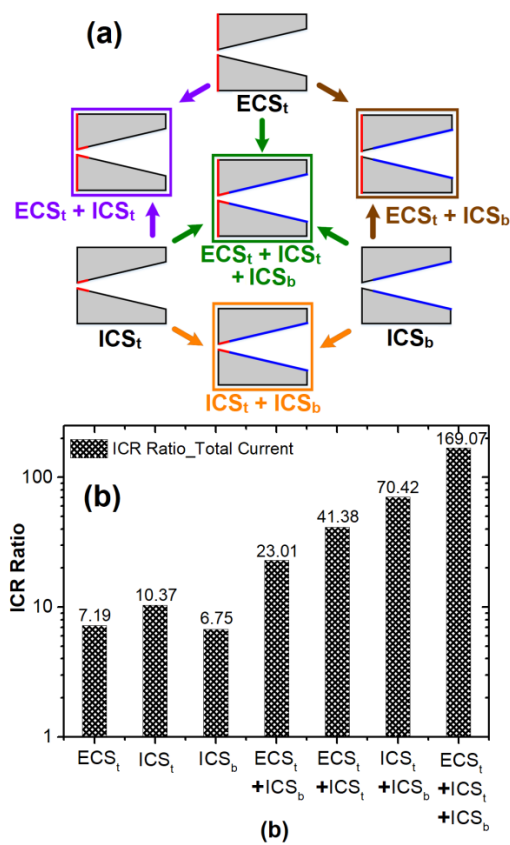

Figure 8 (a) Scheme of simulation models with various combinations of the differently charged inner surface and exterior surface on the tip side. Negative and positive surface charges are shown in red and blue, respectively. (b) Corresponding ICR ratios of current at ±1 V in different simulation models.

As shown in Figure 8, ICR ratios induced by individual charged parts are additive directly, due to their additive effect on the ion concentration regulation along the pore axis (Figure S24). Our results show that with two or more individual charged parts, the ICR ratio can be enhanced from ~20 to ~170. Individual IV curves are shown in Figure S25. We think this finding may provide an effective way to modulate ICR in ultra-short conical nanopores. We would like to emphasize that in order to realize the addition of ICR caused by the charged inner surface at the tip region and that near the base end, the cases with bipolar charge distribution were considered here. While ICR in bipolar nanopores has been well studied, which is caused by another ICR mechanism i.e. the intra-pore ion enrichment and depletion due to different ionic selectivity to cations and anions at oppositely charged pore ends.[50, 51] Our results may provide another possible explanation to ICR in bipolar conical nanopores.

The additive property of ICR can also be used to explain the phenomenon in Figure 1, when combined with any other charged surface, the charged exterior surface on the base side has almost no effect on the current characteristic, which is controlled by the charged inner surface ($IECS_b$ case) or exterior surface on tip side (ECS case) totally.

However, the combined charged parts can balance out ICR if they regulate the ion concentration along the pore axis in opposite directions, as observed in the ACS and $IECS_t$ models (Figure 1),[28, 36, 37] which direction is determined by the charged exterior surface on the tip side because of its larger effect than the charged inner surface on the ion concentration modulation. Consequently, by adjusting the relative ICR caused by different charged parts, large magnitudes of rectification degrees can also be achieved. To illustrate this concept, a model system of a silicon nitride conical nanopore was considered; the pore had a uniform surface charge density of −0.005 C/m$^2$.[38] It was shown that surface modification methods, i.e. attaching different chemical groups to nanopore surfaces, allowed one to control the charge density.[32, 49, 52] For example, treating membrane surface with triethoxysilylpropylmaleamic acid[52] increases the surface charge density to −0.08 C/m$^2$ or even higher. When such highly charged surface becomes the exterior conical pore surface on the tip side,[32] it can induce significant ICR. As shown in Figure S26, when the surface charge density of the exterior surface on the tip side is −0.08 C/m$^2$, and that of the other pore walls is −0.005 C/m$^2$, an ICR ratio as high as ~7.4 is obtained due to the relative larger effect of the charged exterior surface on the tip side on the current rectification.

**Conclusions:**

Our numerical simulations revealed various mechanisms that can lead to ionic current rectification in conical nanopores with 100 nm in length. The weak rectifying properties of short unfirmly charged conical nanopores result from the balancing out of current rectifying caused by the charged inner surface and exterior pore surface on the tip side. Both charged surfaces can induce ICR individually but with opposite directions because of ICP at the tip region and ion enrichment/depletion inside the nanopore, respectively. Due to the stronger influence of ion enrichment/depletion inside the nanopore, the charged exterior surface at the tip end controls the final ICR direction, which has a minimum effective area within ~40 nm away from the tip boundary. While the charged depth of inner surface has a complex influence on ICR. ICR in both directions can be obtained. The ICR direction is determined by the interplay of ICP and intra-pore ion enrichment/depletion under different voltage biases, and can be tuned by bulk salt concentrations. Through the additivity of ICR ratios induced by different charged parts of the conical nanopore, a series of ICR ratios from ~2 to ~170 can be achieved by controlling combinations of differently charged inner surface and exterior surface on the tip side. Based on the advantages of short nanopores, including large ionic permeation coefficient, and high sensitivity to changes in the charge properties and the presence of an analyte, our finding in this work will instruct the design and modification of short conical nanopores in nanofluidic sensors.

**Supporting Information:**

- Simulation details, ionic behaviors and ion concentration distributions in different simulation models, fluid flow characteristics, COMSOL report example.

**Acknowledgments:**

This research was supported by the Qilu Talented Young Scholar Program of Shandong University, Key Laboratory of High-efficiency and Clean Mechanical Manufacture at Shandong University, Ministry of Education, Open Foundation of Advanced Medical Research Institute of Shandong University (Grant No.22480082038411), Natural Science Foundation of Jiangsu Province (BK2020040509), and Guangdong Basic and Applied

**TOC:**

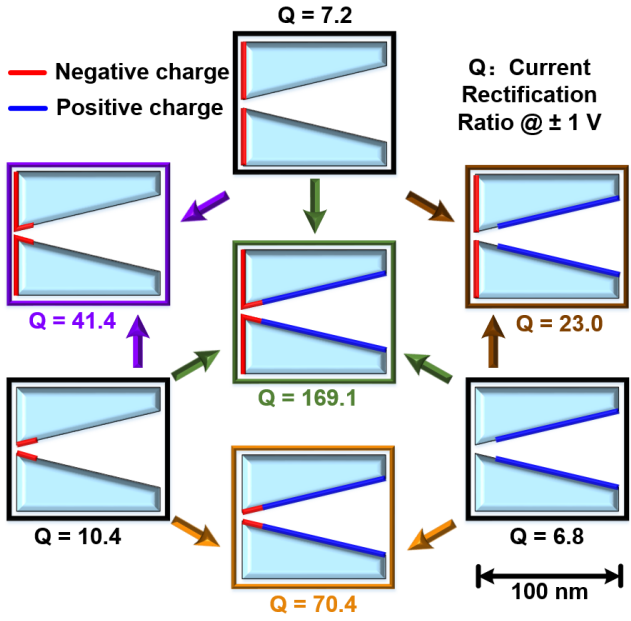